%% file: main.tex
\RequirePackage{latexml}
\iflatexml
\documentclass{article}
\else
\documentclass{bluepreprint}
\fi
\usepackage[T1]{fontenc}
\usepackage{xcolor}
\iflatexml
\newenvironment{frontmatter}{}{}
\newenvironment{keywords}{\par\noindent\textbf{Keywords}\quad}{\par}
\newcommand{\codeandfirmware}[1]{\par\noindent\textbf{Code and firmware}\quad\url{#1}\par}
\newcommand{\bstctlcite}[1]{}
\fi
\usepackage{cite,amsmath,amssymb,graphicx,booktabs}
\usepackage[hidelinks]{hyperref}
\hypersetup{pdftitle={A Four-Target Radar Simulator with Fractional Delay on a PlutoSDR},pdfauthor={Florian Euchner and Stephan ten Brink}}
\iflatexml
\newcommand{\orcidicon}[1]{\href{https://orcid.org/#1}{\textsf{iD}}}
\else
\usepackage{tikz,scalerel}
\usetikzlibrary{arrows.meta,positioning,svg.path}
\definecolor{orcidlogocol}{HTML}{A6CE39}
\tikzset{
  orcidlogo/.pic={
    \fill[orcidlogocol] svg{M256,128c0,70.7-57.3,128-128,128C57.3,256,0,198.7,0,128C0,57.3,57.3,0,128,0C198.7,0,256,57.3,256,128z};
    \fill[white] svg{M86.3,186.2H70.9V79.1h15.4v48.4V186.2z}
                 svg{M108.9,79.1h41.6c39.6,0,57,28.3,57,53.6c0,27.5-21.5,53.6-56.8,53.6h-41.8V79.1z M124.3,172.4h24.5c34.9,0,42.9-26.5,42.9-39.7c0-21.5-13.7-39.7-43.7-39.7h-23.7V172.4z}
                 svg{M88.7,56.8c0,5.5-4.5,10.1-10.1,10.1c-5.6,0-10.1-4.6-10.1-10.1c0-5.6,4.5-10.1,10.1-10.1C84.2,46.7,88.7,51.3,88.7,56.8z};
  }
}

\newcommand\orcidicon[1]{\href{https://orcid.org/#1}{\mbox{\scalerel*{
\begin{tikzpicture}[yscale=-1,transform shape]
\pic{orcidlogo};
\end{tikzpicture}
}{|}}}}

\fi
\newcommand{\orcid}[1]{\textsuperscript{\orcidicon{#1}}}
\title{A Four-Branch Radar Target Simulator\\with Fractional Delay on a PlutoSDR}
\author{Florian Euchner\orcid{0000-0002-8090-1188}\and
Stephan ten Brink\orcid{0000-0003-1502-2571}\thanks{Institute of Telecommunications, University of Stuttgart, 70569 Stuttgart, Germany\\email: \href{mailto:euchner@inue.uni-stuttgart.de,tenbrink@inue.uni-stuttgart.de}{\{euchner,tenbrink\}@inue.uni-stuttgart.de}}}
\date{}
\begin{document}
\bstctlcite{paper-bib-style}
\raggedbottom
\markboth{Preprint}{Euchner and ten Brink}
\begin{frontmatter}
\maketitle
\codeandfirmware{https://github.com/ESPARGOS/plutosdr-radar-target-simulator}
\begin{abstract}
Repeatable test setups support the development and evaluation of integrated sensing and communications (ISAC) hardware and algorithms.
Commercial radar target simulators are expensive, while sub-6~GHz ISAC experiments often need only tens of megahertz of bandwidth within their available spectrum allocation. This makes simpler, low-cost test hardware attractive.
We present a low-cost, open-source radar target simulator based on an ADALM-Pluto without hardware modification, supporting four independently configurable targets and analog bandwidths up to 56~MHz.
The FPGA generates echoes with independent delays, Doppler shifts and gains, using fractional-delay filters for sub-sample adjustment.
For motion-consistent echoes, delay and carrier phase are updated together to reproduce both changing range and frequency-dependent Doppler shifts.
Over-the-air measurements demonstrate simultaneous echoes, sub-sample delay adjustment, and smooth target motion between sample positions.
The FPGA source, firmware and on-device web interface are publicly available.
\end{abstract}
\end{frontmatter}

\section{Introduction}
Radar target simulators generate virtual radar echoes in hardware by delaying and frequency-shifting a received waveform and adjusting its amplitude.
Targets with specified ranges, radial velocities and echo strengths can thus be emulated without moving physical reflectors.
Such setups allow integrated sensing and communications (ISAC) algorithms to be evaluated with communication waveforms and actual radio hardware~\cite{liu}.
For example, Bauhofer \emph{et al.}~\cite{bauhofer2026} use a target emulator to evaluate multi-target tracking with 5G-compliant ISAC radios in a factory environment.

Commercial instruments such as the Rohde \& Schwarz AREG800A~\cite{areg} and Keysight ZA0500A~\cite{keysight} control echo delay, Doppler shift and strength for automotive radar testing.
Their high carrier frequencies and wide bandwidths serve automotive radar, but are not required for many sub-6~GHz ISAC experiments. Here, the available communication spectrum often constrains the sensing bandwidth anyway~\cite{liu}. For allocations of a few tens of megahertz, a simulator need not reproduce gigahertz-wide radar signals. The PlutoSDR is therefore well suited to these experiments: its extended tuning range covers sub-6~GHz carriers, and its analog bandwidth reaches 56~MHz.
Cost matters especially for antenna array testing: the ZA0500A specifies one simultaneous target at one angle~\cite{keysight}, whereas several low-cost simulators could generate echoes from different directions to evaluate the radar's ability to resolve targets at different angles.

\begin{figure}[!b]
\centering
\includegraphics[alt={PlutoSDR transceiver, FPGA echo-generation path and ARM configuration processor.},width=\columnwidth]{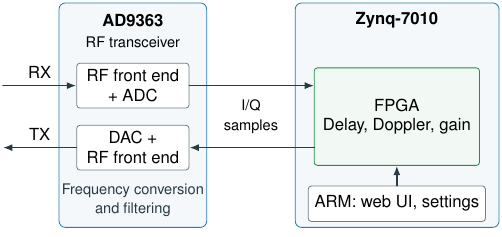}
\caption{PlutoSDR hardware: the FPGA generates echoes from the transceiver's I/Q samples, while the ARM processor handles configuration. Simplified from~\cite{plutohardware}.}
\label{fig:pluto-hardware}
\end{figure}

\begin{figure*}[!t]
\centering
\includegraphics[alt={Radar and simulator setup, with four FPGA paths applying delay, phase and gain.},width=\textwidth]{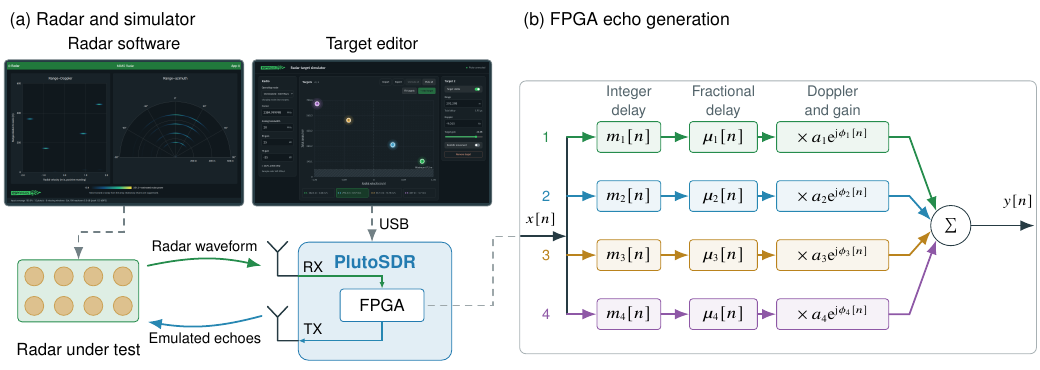}
\caption{(a) Over-the-air setup with radar software and the target editor. (b) Four FPGA target paths, each applying integer and fractional delay, carrier phase and gain. Delay and carrier phase are updated together to generate motion-consistent echoes.}
\label{fig:system}
\end{figure*}

The minimum range of a emulated target is limited by the time required to receive and retransmit the radar signal.
In a monostatic radar, a delay of just $1~\mu$s corresponds to a range of approximately 150~m.
Keeping this delay small requires a short signal path through both the radio transceiver and the digital processing.
We therefore generate the echoes directly in the field-programmable gate array (FPGA) and configure the transceiver to reduce conversion and filter delays.
The PlutoSDR combines an AD9363 RF transceiver with a Zynq-7010 containing FPGA logic and ARM processors, as shown in Fig.~\ref{fig:pluto-hardware}.
Its direct digital RX and TX connections allow the FPGA to generate echoes while the ARM processor handles configuration and the user interface.
The simulator receives and retransmits signals through separate antennas, as shown in Fig.~\ref{fig:system}(a).

Han \emph{et al.}~\cite{han} implemented a radar target simulator on a PlutoSDR by delaying received samples in the FPGA, supporting emulated distances up to 300~km in discrete 5~m steps and introducing Doppler shifts through RF frequency adjustment.
Simultaneous echoes with independent Doppler shifts and minimum latency are not reported, and no source-code repository is provided.

For motion-consistent echoes, the delay needs to change at a rate consistent with the Doppler shift.
Whole-sample delay updates introduce artifacts into the Doppler spectrum~\cite{koerner}.
Prior implementations combine fractional-delay filtering with Doppler correction to emulate moving targets~\cite{diewald2018,diewald2021migration}.
Multirate sampling offers another approach to continuous target motion~\cite{koerner}.

Here, we implement fractional-delay filtering with Doppler correction on an unmodified PlutoSDR for four independently configurable targets.
By updating delay and carrier phase on every input sample, we account for both the changing target range and the dependence of the Doppler shift on signal frequency.
The processing requires no knowledge of the radar waveform or its repetition period.
We evaluate relative delay error, moving-target phase error and self-interference using over-the-air measurements.
All source code and ready-made firmware are publicly available~\cite{repository}.

\section{System model and implementation}
\subsection{Moving targets in monostatic radar}
To describe the emulated echoes, we use monostatic radar geometry, where each echo follows a two-way path from the radar to a point target $k$ and back. At observation time $t$, its range is $R_k(t)$ and its radial velocity is $v_k(t)=\mathrm dR_k(t)/\mathrm dt$, with positive velocity denoting increasing range. For $|v_k|\ll c$, where $c$ is the speed of light, the round-trip delay is
\begin{equation}
 \tau_k(t)=\frac{2R_k(t)}{c}.
 \label{eq:range}
\end{equation}
We assume an echo gain $a_k$ that is constant and frequency independent over the signal band. With delay variable $\tau$, carrier frequency $f_c$ and Dirac impulse $\delta_{\mathrm D}$, this target contributes the passband and baseband channel impulse responses
\begin{align}
 h_{\mathrm{PB},k}(t,\tau)&=a_k\delta_{\mathrm D}(\tau-\tau_k(t)),\\
 h_{\mathrm{BB},k}(t,\tau)&=a_k\mathrm e^{-\mathrm j2\pi f_c\tau_k(t)}
                  \delta_{\mathrm D}(\tau-\tau_k(t)). \label{eq:cir}
\end{align}
Here, $\mathrm j$ denotes the imaginary unit. Thus a transmitted baseband waveform $x(t)$ produces the echo
\begin{equation}
 y_k(t)=a_k x\!\left(t-\frac{2R_k(t)}{c}\right)
             \mathrm e^{-\mathrm j4\pi f_c R_k(t)/c}.
 \label{eq:path}
\end{equation}
Motion changes both the waveform delay and the carrier phase.
For constant radial velocity, $R_k(t)=R_k(0)+v_k t$, and the carrier Doppler shift is
\begin{equation}
 f_{\mathrm D,k}=-f_c\frac{\mathrm d\tau_k(t)}{\mathrm dt}
               =-\frac{2f_c v_k}{c}. \label{eq:velocity}
\end{equation}
To generate motion-consistent echoes, the simulator must reproduce both the changing propagation delay and the carrier phase in~\eqref{eq:path}.
We therefore vary the delay of the baseband samples and apply a phase rotation corresponding to the carrier Doppler shift $f_{\mathrm D,k}$.
The changing delay also shifts a baseband component at frequency $f$ by $-2fv_k/c$. Together with the carrier contribution, this gives the Doppler shift $-2(f_c+f)v_k/c$ expected at its RF frequency $f_c+f$.
Applying only a frequency shift at fixed delay would omit this frequency-dependent contribution and leave the target at a constant range.

In the FPGA, the received waveform is represented by discrete samples $x[n]$ at sample rate $f_s$, where $n$ is the sample index.
A delay line implemented in memory delays the waveform by an integer number of sample periods. Changing this number adjusts the delay in whole-sample steps, but a moving target's propagation delay also takes values between these steps. We therefore use FIR interpolation to implement the fractional part of the delay.

Signal conversion and filtering in the transceiver, transfer across the digital RX/TX interfaces, and the FPGA processing pipeline introduce a fixed delay $\tau_0$, even when no additional delay is configured.
To represent targets at greater ranges, the FPGA adds a programmable delay $d_k[n]$, measured in samples.
The total delay at sample $n$ is therefore $\tau_k(n/f_s)=\tau_0+d_k[n]/f_s$.
We split $d_k[n]$ into an integer part $m_k[n]=\lfloor d_k[n]\rfloor$ and a fractional part $\mu_k[n]=d_k[n]-m_k[n]$, with $0\leq\mu_k[n]<1$.
The integer part determines the effective length of the delay line, while a FIR interpolation filter applies the fractional part.
Each resulting signal $z_k[n]$ is then scaled by the target gain $a_k$ and rotated by the carrier phase $\phi_k[n]$. Summing the four target contributions gives the samples retransmitted by the simulator:
\begin{equation}
 y[n]=\sum_{k=1}^{4}a_k z_k[n]\mathrm e^{\mathrm j\phi_k[n]}. \label{eq:model}
\end{equation}
To follow the motion model in~\eqref{eq:path}, the carrier phase must satisfy $\phi_k[n]=-2\pi f_c\tau_k(n/f_s)$, up to a constant initial phase.
For a target moving at constant radial velocity, we therefore update both its delay and its phase on every input sample:
\begin{align}
 d_k[n+1]&=d_k[n]+\frac{2v_k}{c}=d_k[n]-\frac{f_{\mathrm D,k}}{f_c}, \label{eq:motion}\\
 \phi_k[n+1]&=\phi_k[n]+2\pi\frac{f_{\mathrm D,k}}{f_s}\pmod{2\pi}. \label{eq:phase}
\end{align}
The delay increment is expressed in samples per input sample, so the sample rate cancels in~\eqref{eq:motion}.
The simulator also supports fixed-delay echoes with independently configured Doppler shifts. In this mode, only the carrier phase is advanced, while the delay is held constant.

The user interface displays equivalent range and radial velocity as
$R_k[n]=c(\tau_0+d_k[n]/f_s)/2$ and
$v_k=-cf_{\mathrm D,k}/(2f_c)$. The range includes both hardware and programmable delay. In an over-the-air setup, propagation from the radar to the simulator and back can add an appreciable delay.

\begin{figure}[!t]
\centering
\includegraphics[alt={Integer delay followed by a six-tap fractional-delay filter using a coefficient lookup table.},width=\columnwidth]{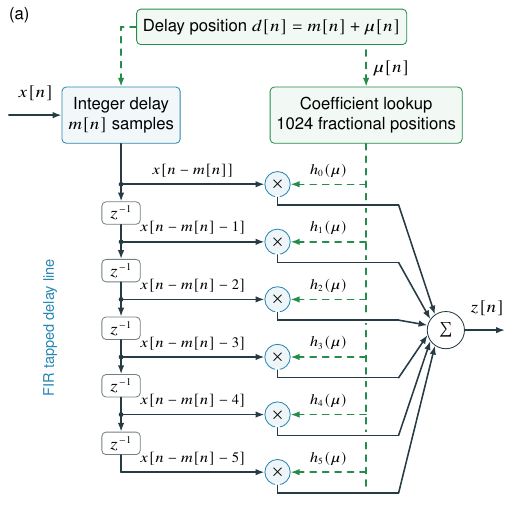}
\par\vspace{1mm}
\includegraphics[alt={Six interpolation coefficients at a fractional delay of approximately 0.6 samples, compared with ideal sinc interpolation.},width=\columnwidth]{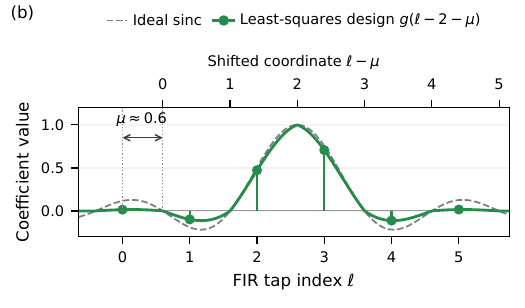}
\caption{Six-tap delay processing. (a) Block diagram of the per-target delay implementation. (b) Six filter coefficients for fractional delay $\mu=614/1024\approx0.6$, obtained from the quantized solutions of~\eqref{eq:filter-design}. Each coefficient set is optimized separately, so smooth joins between curve segments are not imposed.}
\label{fig:fractional}
\end{figure}

\subsection{FPGA implementation}
The PlutoSDR's FPGA has limited on-chip memory, logic and hardware multipliers, requiring a compromise between the number of simultaneous targets, the maximum programmable delay and the interpolation filter length.
Four target paths are supported, with a maximum programmable delay of 12276 radio samples and interpolation filters of length $L=6$ or $L=8$ depending on the operating mode.
The four target paths are implemented in the Zynq-7010 FPGA.
The input samples are stored in FPGA memory. The requested integer delay determines the memory address from which samples are read.
For direct interpolation at the radio sample rate, a fractional-delay filter~\cite{laakso} with $L$ taps computes
\begin{equation}
\begin{aligned}
 z_k[n]&=\sum_{\ell=0}^{L-1}h_\ell(\mu_k[n])\,x[n-m_k[n]-\ell]\\
 &=\sum_{\ell=0}^{L-1}g(\ell-p-\mu_k[n])\,x[n-m_k[n]-\ell].
\end{aligned}
\end{equation}
Here, $h_\ell(\mu)$ is the coefficient of tap $\ell$ at fractional position $\mu$. The function $g$ describes the filter coefficients across all fractional delays. For a given delay $\mu$, the coefficients are obtained by sampling the shifted function at the integer tap positions.
The interpolation introduces a fixed offset $p=L/2-1$ samples, placing taps on both sides of the fractional interpolation position. The desired filter delay is therefore $p+\mu$ samples. The fixed contribution $p$ is included in $\tau_0$.
An ideal full-band fractional-delay filter has the infinitely long impulse response $\operatorname{sinc}(\ell-p-\mu)$, with $\operatorname{sinc}(u)=\sin(\pi u)/(\pi u)$. Truncating this sinc does not minimize the error within the simulator's signal band. Alternatives include windowed-sinc filters, Lagrange interpolation with a maximally flat response near DC, and least-squares approximation~\cite{laakso}. We choose least squares to fit the required bandwidth with the available taps. For each fractional position $\mu$, the real coefficients $h_\ell=h_\ell(\mu)$ are precomputed with equal weighting across the maximum supported bandwidth $B$ and unity DC gain. With filter input sample rate $f_{\mathrm i}$, the optimization is
\begin{equation}
\begin{aligned}
 \underset{h_0,\ldots,h_{L-1}\in\mathbb R}{\operatorname{minimize}}\quad &
 \int_{-B/2}^{B/2}\left|
 \sum_{\ell=0}^{L-1}h_\ell\mathrm e^{-\mathrm j2\pi f\ell/f_{\mathrm i}}
 -\mathrm e^{-\mathrm j2\pi f(p+\mu)/f_{\mathrm i}}\right|^2\,\mathrm df\\
 \text{subject to}\quad & \sum_{\ell=0}^{L-1}h_\ell=1.
\end{aligned}
\label{eq:filter-design}
\end{equation}
The coefficients are precomputed for 1024 fractional delays. These solutions define $g$ through $g(\ell-p-\mu)=h_\ell(\mu)$, as illustrated in Fig.~\ref{fig:fractional}(b). The approximation errors are evaluated in Section~\ref{sec:fractional-accuracy}.

Coefficient storage and arithmetic are shared between target paths. Multiplications use dedicated hardware multipliers and additional FPGA logic.

For Doppler and gain control, each target's phase advances according to~\eqref{eq:phase}. Gain-scaled sine and cosine values are read from a per-target table and used to multiply the delayed signal by $a_k\mathrm e^{\mathrm j\phi_k[n]}$.
When settings change, a second set of tables is filled from a shared sine/cosine reference and activated once all four are complete. This double buffering allows uninterrupted echo generation, while scaling the table entries avoids a separate gain multiplication for every sample.

For moving targets, the delay advances on the same sample clock according to~\eqref{eq:motion}.
Targets are modeled as point reflectors with constant gains, without range-dependent attenuation or micro-Doppler from extended objects.
Since all echoes leave the same antenna, they have the same apparent direction, which does not change with the programmed motion.

FPGA resource use is summarized in Table~\ref{tab:profiles}. Coefficients and signal samples use fixed-point representation, introducing quantization error. Output clipping is detected in the FPGA and reported through the web interface.
The ARM processor serves a web interface for configuring the simulator. The PlutoSDR presents a network interface to the host computer over USB, allowing this interface to be accessed in a web browser.

\input{figures/profiles.tex}

\section{Transceiver configuration}
\label{sec:radio}
\subsection{Standard and overclocked modes}
Reducing receive-to-transmit latency allows the simulator to represent closer targets.
In the PlutoSDR, the AD9363 and FPGA are connected by separate 12-bit parallel data buses for RX and TX~\cite{plutohdl}. I and Q are transferred on opposite clock edges, giving one complex sample per clock cycle in each direction. In \emph{standard mode}, the interface runs at 61.44~MSa/s with the receive interface clock DATA\_CLK at its specified 61.44~MHz limit~\cite{ad9363}. The AD9363's internal decimation and interpolation filters remain enabled, while its programmable FIR filters are disabled.

To reduce this delay, we raise the sample rate to 160~MSa/s and the interface clock to 160~MHz, and bypass selected receive and transmit filter stages. We call this configuration \emph{overclocked mode}, since its interface clock exceeds the specified limit.
The higher rate reduces the minimum range but also the maximum delay available from a fixed-size sample memory. Table~\ref{tab:profiles} compares the settings and performance of both modes.

The operating mode can be selected through the web interface. During startup, RX and TX operation is checked using pseudorandom sequences with the target paths active. If the check fails in overclocked mode, standard mode is attempted. The targets remain muted if neither mode passes.

\subsection{Tuning range, bandwidth and delay}
The AD9363 is specified for 325--3800~MHz carriers and 20~MHz analog bandwidth. The tuning-range extension documented by Analog Devices~\cite{plutoextension} configures the driver to treat it as an AD9364, permitting 70~MHz--6~GHz tuning and up to 56~MHz analog bandwidth without hardware modification. These settings exceed the AD9363 specification. The extension applies independently of standard or overclocked operation, with a common RX/TX bandwidth setting.
The hardware delays at 56~MHz bandwidth are listed in Table~\ref{tab:profiles}. Dedicated ADC--FPGA--DAC hardware can achieve lower latency, with 73.3~ns reported by Sobotka and Adler~\cite{sobotka2026latency}. Their result excludes the RF front end, whereas our delay includes the AD9363 transceiver.
The FIR coefficients used to implement fractional target delays are fitted over $\pm28$~MHz in both modes. In overclocked mode, $L=6$ taps operate directly at $f_{\mathrm i}=f_s=160$~MSa/s.
At 61.44~MSa/s, a signal with bandwidth 56~MHz occupies almost the full digitally representable band. Applying a longer fractional-delay filter directly at this rate would repeat its computation and coefficient storage for each target. Instead, a shared 95-tap filter first doubles the sample rate to $f_{\mathrm i}=122.88$~MSa/s, followed by an $L=8$ fractional-delay filter per target. The shared filter is a half-band low-pass design whose symmetry and alternating zero coefficients reduce the multiplication count.

At the upsampled sampling rate, the same physical delay corresponds to $2d_k[n]$ samples. Its integer and fractional parts select the stored samples and interpolation coefficients. Every second filtered output is kept to return to 61.44~MSa/s, and only these outputs are calculated in the FPGA. Both filters' fixed delays are included in $\tau_0$. Each mode has a separate coefficient table covering all supported fractional delays.

\begin{figure}[!t]
\centering
\includegraphics[alt={ESPARGOS radar array above the PlutoSDR simulator with separate receive and transmit antennas.},width=.52\columnwidth]{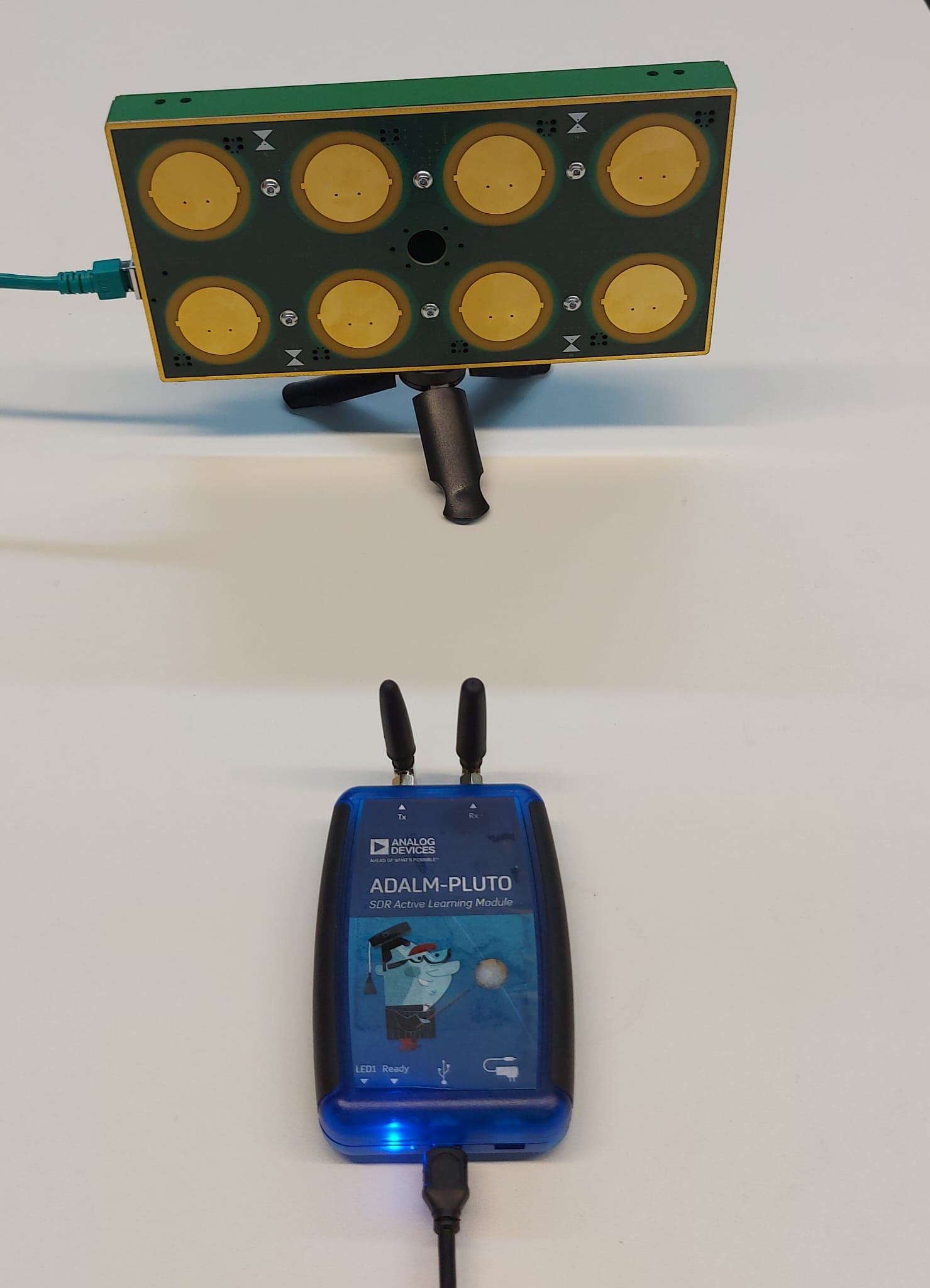}
\caption{Over-the-air setup: ESPARGOS radar array (top) and PlutoSDR target simulator with separate RX and TX antennas (bottom).}
\label{fig:experimental-setup}
\end{figure}

\begin{figure}[!t]
\centering
\includegraphics[alt={Measured range-Doppler map resolving four programmed targets at different ranges and Doppler shifts.},width=\columnwidth]{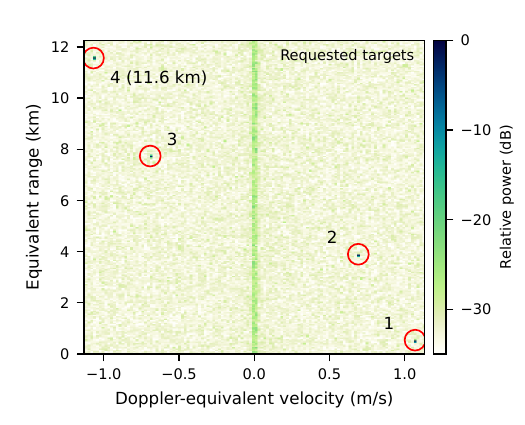}
\caption{Four fixed-delay targets with independently configured Doppler, measured over the air at 2385~MHz. Requested targets are circled. Colors show $10\log_{10}(P/P_{\mathrm{ref}})$, where $P$ is the measured power and $P_{\mathrm{ref}}$ is the peak power of the strongest echo after coherent receive-channel combining.}
\label{fig:results}
\end{figure}

\section{Evaluation}
\subsection{Fixed-delay targets with Doppler}
An ESPARGOS array~\cite{espargos} is used as the radar. It supports several radar modes, including pulse compression and orthogonal frequency-division multiplexing (OFDM) radar. OFDM radar is used for the measurements presented here.
One antenna in the array transmits at 2385~MHz, and the signals from the other seven antennas are recorded. The PlutoSDR is placed in front of the array, using separate RX and TX antennas, as shown in Fig.~\ref{fig:experimental-setup}.
The channel response on each OFDM subcarrier is estimated from the received signal and the known transmitted symbol. The radar's self-interference signal, caused by coupling between its transmit and receive antennas, is recorded separately with the simulator muted. This signal serves as the phase reference for estimating echo delays. Stationary self-interference and clutter are suppressed.

To demonstrate simultaneous targets across the available delay range, four echoes are configured in overclocked mode with added delays of 3.20, 25.59, 51.19 and 76.73~$\mu$s. These correspond to equivalent monostatic ranges of approximately 0.54, 3.90, 7.74 and 11.56~km.
Their Doppler shifts are $-17$, $-11$, $+11$ and $+17$~Hz, corresponding to radial speeds of about 0.7 and 1.1~m/s.
For Fig.~\ref{fig:results}, the radar samples at 10~MSa/s and uses 1024-sample windows with 800 occupied subcarriers. The seven receive channels are combined coherently. A 120~s recording resolves all four echoes, including the farthest at about 11.6~km.

The minimum receive-to-transmit delays in Table~\ref{tab:profiles} are measured with zero added delay and $+17$~Hz Doppler, separating the echo from stationary interference. The 56~MHz bandwidth setting is evaluated with OFDM recordings at 80~MSa/s.

\subsection{Fractional-delay accuracy}
\label{sec:fractional-accuracy}
The interpolation filters are evaluated numerically against an ideal delay with unit gain. For each of the 1024 fractional-delay settings, the absolute difference between the implemented and ideal complex frequency responses is calculated on a dense frequency grid spanning $\pm28$~MHz. The largest difference over all evaluated frequencies and settings, expressed as $20\log_{10}|H-H_{\mathrm{ideal}}|$, is $-60.6$~dB in standard mode and $-61.8$~dB in overclocked mode. The standard-mode calculation includes the shared upsampling filter and both sampling phases. Both calculations include coefficient quantization.

To evaluate sub-sample delay control, a single emulated target is measured using an OFDM signal with 55~MHz bandwidth. Its $+17$~Hz Doppler separates the echo from stationary self-interference and clutter. Relative delay is estimated by fitting the phase change across OFDM subcarriers to the linear phase response of a single delayed echo.

\begin{figure}[t]
\centering
\includegraphics[alt={Measured versus requested small range changes in standard and overclocked modes.},width=\columnwidth]{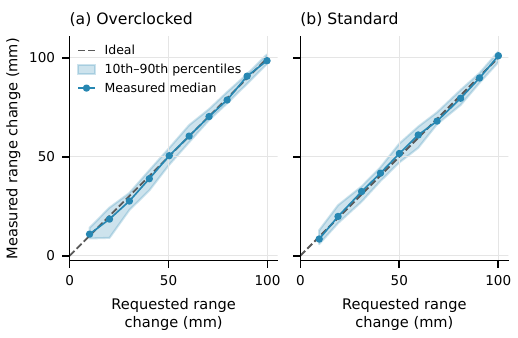}
\caption{Relative-delay measurements expressed as equivalent monostatic range changes, $\Delta R=c\Delta\tau/2$, at 56~MHz bandwidth. At each requested range, markers show the median and shading the 10th--90th percentiles of ten measurements from independent sweeps. The dashed line denotes exact agreement.}
\label{fig:range-steps}
\end{figure}

Figure~\ref{fig:range-steps} shows ten sweeps per mode over approximately 100~mm in equivalent monostatic range, in nominal 10~mm increments. Range changes are measured relative to a reference with no programmable delay added beyond the simulator's hardware latency. This latency corresponds to equivalent monostatic ranges of 62.1~m in overclocked mode and 197.7~m in standard mode. Propagation between the radar and simulator adds a small fixed offset to these reference ranges. Each sweep uses a separate RF calibration, 4~s per position and initial and final reference recordings to correct linear drift.

The largest absolute median errors are 2.6~mm in overclocked mode and 1.6~mm in standard mode. Individual errors reach 21~mm and 12~mm, respectively. Additional sweeps across a complete sample interval confirmed that the requested delay changes were reproduced.

\subsection{Frequency-dependent Doppler}
To verify that emulated motion reproduces the Doppler shift across the signal band, the Doppler shift is measured independently on each OFDM subcarrier. Ten 20~s recordings are acquired for each of two settings in overclocked mode: motion-consistent echoes and fixed-delay echoes with Doppler shifts. Both use an initial added delay of 6.25~$\mu$s, approximately $+40$~Hz carrier Doppler and 176 occupied subcarriers spanning 55~MHz. The two settings are interleaved between recordings.

\begin{figure}[t]
\centering
\includegraphics[alt={Measured subcarrier Doppler and its difference from theory for moving and fixed-delay targets.},width=\columnwidth]{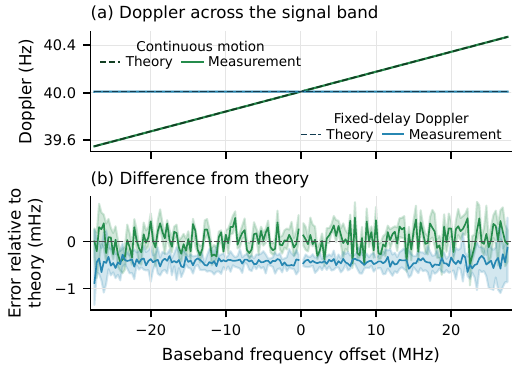}
\caption{Doppler versus OFDM subcarrier frequency at a 2385~MHz carrier. Lines show measured medians and shading the 10th--90th percentiles of ten independent recordings per setting. (a) Measured Doppler, with theoretical predictions shown dashed. (b) Difference from the respective prediction.}
\label{fig:ofdm-doppler}
\end{figure}

Figure~\ref{fig:ofdm-doppler} shows the measured Doppler on each subcarrier. Motion-consistent echoes produce the expected frequency dependence $f_{\mathrm D}f/f_c$, whereas Doppler at fixed delay is nearly independent of baseband frequency. Figure~\ref{fig:ofdm-doppler}(b) shows small differences between measured and theoretical Doppler across the signal band.

\begin{figure}[t]
\centering
\includegraphics[alt={Measured range and velocity of two moving targets, with settings and predicted feedback tracks.},width=\columnwidth]{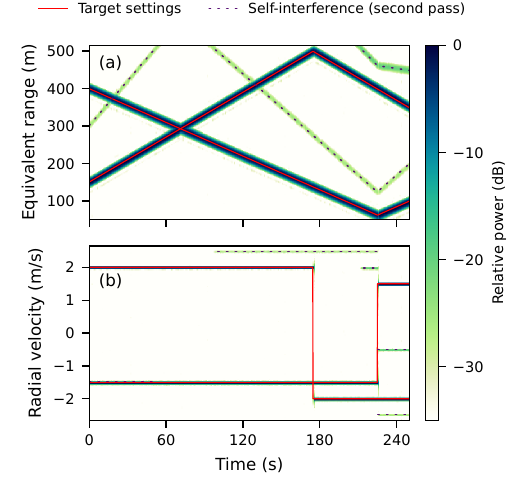}
\caption{Two moving targets: (a) range and (b) radial velocity. Solid lines show target settings and dotted lines predicted self-interference, including Doppler aliasing. From each two-second range--Doppler map, (a) shows the strongest echo power at each range across all Doppler values, and (b) the strongest power at each velocity within 60--500~m. Both use the same normalization to the strongest echo in this range over the recording.}
\label{fig:trajectory}
\end{figure}

\subsection{Continuous motion of multiple targets}
To demonstrate simultaneous target motion, two targets are observed over 250~s using OFDM radar with 18.75~MHz occupied bandwidth. Figure~\ref{fig:trajectory} shows their measured ranges and radial velocities, starting near 150 and 400~m at $+2$ and $-1.5$~m/s. The range tracks cross while the Doppler velocities remain distinct. At each direction reversal, the range slope and Doppler sign change together. These observations show that the two targets follow their configured trajectories independently, with consistent range and Doppler throughout the recording.

\subsection{Self-interference}
\begin{figure}[t]
\centering
\includegraphics[alt={Transmit-to-receive antenna coupling sends emulated echoes through the target paths again.},width=\columnwidth]{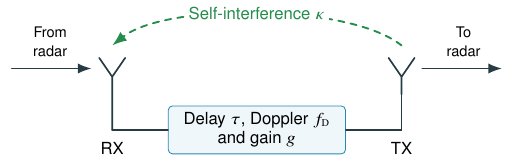}
\caption{Self-interference through TX-to-RX antenna coupling causes echoes to pass through the simulator again.}
\label{fig:feedback}
\end{figure}

Several weaker tracks in Fig.~\ref{fig:trajectory} correspond to self-interference. Part of the transmitted echo is coupled back into the simulator's receive antenna and processed a second time, as illustrated in Fig.~\ref{fig:feedback}. For target indices $i$ and $j$ (including $i=j$), this produces an additional echo near range $R_i+R_j$ with Doppler shift $f_{\mathrm D,i}+f_{\mathrm D,j}$, plus the short antenna-to-antenna propagation delay. The dotted lines in Fig.~\ref{fig:trajectory} show these predictions, including Doppler aliasing, and follow the observed weaker tracks. Improved isolation between the simulator's antennas or lower transmit gain reduces these echoes.

\section{Conclusion}
The results show that a low-cost PlutoSDR can generate repeatable virtual echoes for sub-6~GHz ISAC experiments with bandwidths up to 56~MHz. Four targets can be controlled independently without hardware modification. Fractional-delay processing provides fine range adjustment, while coordinated delay and phase updates reproduce both range evolution and the frequency dependence of Doppler. The observed self-interference echoes highlight the importance of antenna isolation in over-the-air setups.

The open-source FPGA design, firmware and browser interface allow the simulator to be adapted to different experiments~\cite{repository}. A useful next step is coordinated operation of several simulators placed at different angles, enabling repeatable tests of angular separation together with range and Doppler estimation.

\section*{Acknowledgment}
\footnotesize
This work is supported by the German Federal Ministry of Research, Technology and Space (BMFTR) within the SENSATION project (grant no. 16KIS2532).
OpenAI Codex assisted software development, manuscript drafting and checks for errors.
\normalsize

\bibliographystyle{IEEEtran}
\bibliography{references}
\end{document}

%% file: figures/profiles.tex
\begin{table}[!t]
\caption{Technical specifications in standard and overclocked modes}
\label{tab:profiles}
\centering
\renewcommand{\arraystretch}{1.16}
\setlength{\tabcolsep}{3pt}
\begin{tabular}{@{}lrr@{}}
\toprule
 & Standard & Overclocked\\
\midrule
Sample rate (MSa/s), DATA\_CLK (MHz) & 61.44 & 160\\
RX/TX analog bandwidth (MHz) & 0.2--56 & 0.2--56\\
Fractional-delay filter taps & 8 & 6\\
Shared upsampling filter taps & 95 & None\\
RF delay at widest setting & 1.319~$\mu$s & 0.414~$\mu$s\\
Minimum range at widest setting & 197.7~m & 62.1~m\\
Range adjustment step & 2.38~mm & 0.915~mm\\
Maximum equivalent range & 30.1~km & 11.6~km\\
Doppler adjustment step & 0.0143~Hz & 0.0373~Hz\\
\midrule
Logic LUTs (of 17600) & 12196 & 14989\\
Block RAM tiles (of 60) & 46 & 57\\
DSP slices (of 80) & 77 & 78\\
\bottomrule
\end{tabular}
\end{table}